# From Community Network to Community Data: Towards Combining Data Pool and Data Cooperative for Data Justice in Rural Areas

Jean Louis Fendji Kedieng Ebongue

University of Ngaoundere, Ngaoundere, Cameroon

Stellenbosch Institute for Advanced Study, Stellenbosch, South Africa

Email: lfendji@gmail.com

**Abstract** :

This study explores the shift from community networks (CNs) to community data in rural areas, focusing on combining data pools and data cooperatives to achieve data justice and foster and a just AI ecosystem. With 2.7 billion people still offline, especially in the Global South, addressing data justice is critical. While discussions related to data justice have evolved to include economic dimensions, rural areas still struggle with the challenge of being adequately represented in the datasets. This study investigates a Community Data Model (CDM) that integrates the simplicity of data pools with the structured organization of data cooperatives to generate local data for AI for good. CDM leverages CNs, which have proven effective in promoting digital inclusion, to establish a centralized data repository, ensuring accessibility through open data principles. The model emphasizes community needs, prioritizing local knowledge, education, and traditional practices, with an iterative approach starting from pilot projects. Capacity building is a core component of digital literacy training and partnership with educational institutions and NGOs. The legal and regulatory dimension ensures compliance with data privacy laws. By empowering rural communities to control and manage their data, the CDM fosters equitable access and participation and sustains local identity and knowledge. This approach can mitigate the challenges of data creation in rural areas and enhance data justice. CDM can contribute to AI by improving data quality and relevance, enabling rural areas to benefit from AI advancements.

## 1 Introduction

Discussions on data justice have evolved to address extreme imbalances in economic opportunities; however, the reality in the Global South is different. Big Tech companies still struggle to penetrate, mainly because of their low Internet penetration rate. According to ITU's figures, 2.7 billion people are still offline (ITU, 2020). Only 36% of the population lives online in the least developed and landlocked developing countries. These figures emphasize the need to shift debates to the Global South. Efforts are being made by some organizations, including the Global Partnership on AI and the Alan Turing Institute, to relocate the discussion on data justice to the Global South (Leslie, Katell, Aitken, Singh, Briggs, Powell, Rincon, et al., 2022). However, the term "Global South" seems overly broad and overlooks an endangered segment: rural areas that are either unconnected or restricted to Level 1 access to the Internet, resulting in poor service quality, low literacy levels, and lack of relevant content. Shifting the debate to this segment introduces new challenges as it is no longer solely about biasing or misusing data. In fact, data do not yet exist in digital form in these areas. There is a need to reconsider early discussions on data justice around the right to not be ignored, erased, or reified. The work in (J. L. K. E. Fendji, 2024) anticipates the impending challenges for the unconnected with the advent of generative AI and the lack of data that convey their realities. This emphasizes the urgency of data creation. How can an environment enable data production in rural areas?

Owing to the declining price of wireless equipment, CNs have emerged as attractive solutions for connecting rural areas (Rey-Moreno, 2014). Typically initiated and sustained by local communities, CNs serve as catalysts for digital inclusion. The creation of local content has been a prominent topic, as highlighted in various Internet Governance Forums since 2014 (IGF, 2020). However, the impact on CNs has been limited thus far. This can be attributed to a lack of strategy and misunderstanding of the significance of sustaining self-identity and preserving cultural heritage in the era of AI. Therefore, there is a pressing need to generate local data and knowledge. Achieving this goal will necessitate innovative approaches that can capitalize on existing infrastructure and overcome challenging constraints, such as a lack of skills.

This study advocates a paradigm shift from CNs to community data in rural areas. It goes beyond the traditional considerations of CNs to address the challenges in local data creation and management. The proposed Community Data Model (CDM) takes inspiration from data sharing pools and data cooperatives commonly employed in small and medium enterprises(Micheli et al., 2020). A data-sharing pool typically involves a collaborative arrangement in which individuals or entities contribute their data to a centralized repository for mutual benefit. On the other hand, a data cooperative is a more structured organization where members collectively own and manage data-related assets and resources. Members participate in decision-making processes and have a say in how data are collected, shared, and utilized. CDM seeks to combine the simplicity of using a common repository with the structured organization of data cooperatives.

The remainder of this paper is organized as follows. Section 2 briefly presents the concept of data justice with a focus on rural areas. Section 3 provides insights into CNs and their importance. The data pool and data cooperative are discussed in Section 4, followed by the CDM in Section 5. Finally, Section 6 depicts how the pillars of data justice are present in the CDM, and Section 7 discusses how the CDM can be leveraged for AI for good in rural areas.

## 2 Data justice in rural areas

### 2.1 Data justice: concepts and pillars

Data justice involves the fair treatment of individuals and communities in the gathering, utilization, and sharing of data (Leslie, Katell, Aitken, Singh, Briggs, Powell, Rincón, et al., 2022). It includes principles that aim to prevent the reinforcement of existing inequalities and promote fairness and inclusivity through data practices. Data justice is essential because data-driven decisions now greatly influence various aspects of life. Ensuring data justice ensures that everyone benefits equally from advances in data without facing biases or exploitation, with a particular focus on marginalized communities.

Recent discussions have delineated six key pillars around which the concept of data justice revolves: power, equity, access, identity participation, and knowledge. Power involves controlling and influencing the data collection, usage, and governance. It is important to balance this power to prevent certain groups from dominating data resources. Equity ensures fair treatment of data practices, particularly for historically marginalized groups. Access aims to provide equal access to data and necessary digital tools, eliminating exclusion due to socioeconomic barriers. Identity focuses on protecting individual and community identities in data representation. Participation enables communities to actively participate in decision-making processes regarding data practice. Knowledge promotes literacy education as a means of empowering communities to engage with available datasets for development needs.

## 2.2 Relevance of data justice in rural communities

Rural areas face specific challenges that highlight the importance of data justice. These regions often lack adequate infrastructure, have limited access to technology, and show low levels of digital literacy, further contributing to their exclusion from the benefits of the digital age. Limited availability of the Internet and restricted access to digital technologies are common issues in rural settings, hindering effective data collection and utilization. In addition, priority should be given to cultural sensitivity by respecting the identities and traditional knowledge of rural communities as a means of preventing marginalization or misrepresentation. Moreover, it is essential for rural communities to have control over how their data are used, empowering them to ensure that they serve their interests, rather than external agendas. More critically, as highlighted by (J. L. K. E. Fendji, 2024), it is essential to recognize the importance of rural communities when creating datasets. Without doing so, AI systems may cause significant harm when interacting with these communities, as their reality is not considered. Therefore, there is a need for a local data creation process that can help produce quality data for a just AI ecosystem.

## 2.3 Data justice and local data creation

The creation and management of local data are crucial for achieving data justice in rural areas. Local data reflects the unique needs, challenges, and aspirations of rural communities. Relevant local data empower communities to make informed decisions tailored to their specific contexts and advocate for their needs more effectively. Involving the community to collect and manage local data ensures correct labelling and accuracy, while fostering ownership and empowerment among members. Leveraging existing connectivity initiatives in rural areas is essential for gathering real-time information that reflects the community's reality.

# 3 Community networks

## 3.1 Definition

Community Networks (CNs) are grassroots, decentralized communication infrastructures developed and maintained by local communities to provide Internet access and digital services, particularly in underserved or remote areas (Supporting the Creation and Scalability of Affordable Access Solutions: Understanding Community Networks in Africa, 2017). The traditional market-driven approach often neglects areas with low population density or limited economic potential, resulting in significant portions of the population being left unconnected. Unlike commercial Internet service providers, CNs are built with the community's specific needs in mind, emphasizing local ownership, participation, and benefit. They originated as a response to the digital divide, which left many rural and marginalized communities without reliable Internet access.

## 3.2 Benefits of CNs

Community Networks (CNs) provide various benefits to both rural and underserved areas. By providing affordable Internet access, CNs help bridge the gap between urban and rural areas. They enable access to essential services, such as healthcare, education, and government resources, which are increasingly migrating to digital platforms. Additionally, CNs provide a platform for local content creation and sharing, allowing the preservation and promotion of local cultures, languages, and traditions. This promotes inclusivity and enables communities to participate actively in the digital space, contributing to a more diverse and representative online environment. Furthermore, CNs foster economic growth by facilitating online entrepreneurship, connecting local businesses to larger markets, and creating opportunities for skill development and employment. The democratization of Internet access through CNs also supports civic engagement and community organization, allowing for better coordination of local initiatives and amplifying voices from traditionally marginalized groups.

## 3.3 Governance structure of CNs

The framework for CN governance can be divided into five primary aspects: organizational structure, decision-making processes, community engagement, sustainability and long-term planning, and legal and regulatory compliance (Micholia et al., 2018). The organizational structure includes a Board of Directors with diverse community leaders, technical experts, and stakeholder representatives supported by an executive team managing daily operations and specialized committees. Effective decision making ensures transparency and fairness and regular meetings with annual general meetings for major decisions. Community engagement is facilitated through communication channels and educational programs, fostering active participation and collective ownership. Sustainability and long-term planning focus on developing sustainable revenue streams and strategic planning with community inputs to ensure stability and alignment with evolving needs. Legal and regulatory compliance involves incorporation and adherence to regulations.

## 3.4 Potential of CNs in data creation

Community Networks (CNs) can play a significant role in data creation in rural communities. Data creation is the process of generating raw information, such as numerical data, images, audio, and text, intended for the analysis and decision-making processes. This process focuses on capturing accurate and relevant data points for computational purposes. By contrast, content creation, which has been presented as the flagship output of leveraging CNs, is mainly about producing media, such as articles, videos, or graphics, designed to engage, inform, or entertain an audience. While data creation is aimed at analysts and decision makers through dataset production, content creation targets a broader audience and often involves creative storytelling.

By providing Internet access and digital services, CNs enable the collection, sharing, and management of not only local knowledge, but also local data. This facilitates the creation of a more comprehensive and accurate representation of a community's needs, challenges, and cultural practices. Moreover, through these networks, individuals can contribute their unique knowledge and experiences, ensuring that the collected data are culturally sensitive and relevant to the community's context. However, the governance structure of CNs must be adapted, taking advantage of well-known systems, such as data pools and data cooperatives.

# 4 Data pools and data cooperatives: concepts and models

## 4.1 Understanding data pools

A data pool is a centralized repository in which multiple entities contribute data, enabling collective access and analysis. Its primary purpose is to aggregate data from various sources and to create a comprehensive dataset for diverse analytical applications. Data pools are prevalent in industries such as healthcare, finance, and retail, where combining data from different sources yields deeper insights and fosters innovation. For example, healthcare providers use centralized health information exchanges (HIEs) to enhance diagnosis and treatment plans by accessing complete patient history (Dash et al., 2019). Financial institutions share transaction data to detect fraud and manage risks more effectively, while retailers pool sales and inventory data to understand market trends, optimize supply chains, and improve customer experiences (Aćimović & Stajić, 2019). The advantages of data pools include creating comprehensive datasets for better decision-making, sharing storage and management costs. However, they also pose challenges, such as data privacy concerns and difficulties in maintaining data quality and standardization(Bormida, 2021).

## 4.2 Understanding data cooperatives

A data cooperative is a collective organization in which individuals or small entities pool their data resources and share the benefits of data aggregation and analysis, emphasizing member ownership and democratic governance (Baars et al., 2021). This model ensures that all participants have a say in how the data are used and managed. Examples include MIDATA Cooperative in Switzerland, where individuals control their health data (Mòdol, 2019); and Farm Hack in the USA, where farmers share agricultural data and innovations(Rotz et al., 2019). Data cooperatives offer significant benefits, such as member control, equitable benefit distribution, enhanced privacy, and strong community building. However, they also face challenges such as complex governance, resource-intensive setup and maintenance, and scalability. Democratic governance requires effective decision-making and conflict resolution mechanisms, while maintaining engagement and control becomes more difficult as cooperatives scale up. Despite these challenges, data cooperatives provide a promising model for ethical and collaborative data management, ensuring that the benefits of data aggregation are shared fairly among members, promoting transparency and fostering a sense of community.

Data pools suit large-scale data needs and efficiency, whereas data cooperatives are ideal for contexts that prioritize member control and community involvement. Both models contribute to improved data practices, which are tailored to the individual objectives and values of users. Furthermore, these models can be combined to unlock the benefits of each approach.

# 5 Community data model: combining data pool and data cooperative

The data cooperative and the data pool can be combined to build the CDM, as shown in Figure 1. The data cooperative provides a community-driven approach that involves community members in data collection, management, and decision-making processes, offering incentives and defining roles and responsibilities. The data pool integrates various sources of raw data within the community. The data pool is fed into a centralized data management platform, providing a diverse range of community-relevant data types. These raw data are then processed, managed, and utilized for AI for Good.

The Community Data Model (CDM) aligns with CNs to reduce additional costs, minimize workload, and facilitate seamless integration. Five key dimensions are considered: Governance and Management

(GM), Infrastructure and Technology (IT), Social Impact and Community Needs (SICN), Capacity Building and Sustainability (CBS), and Legal and Regulatory Environment (LRE). The main points of each CDM dimension are shown in Figure 2.

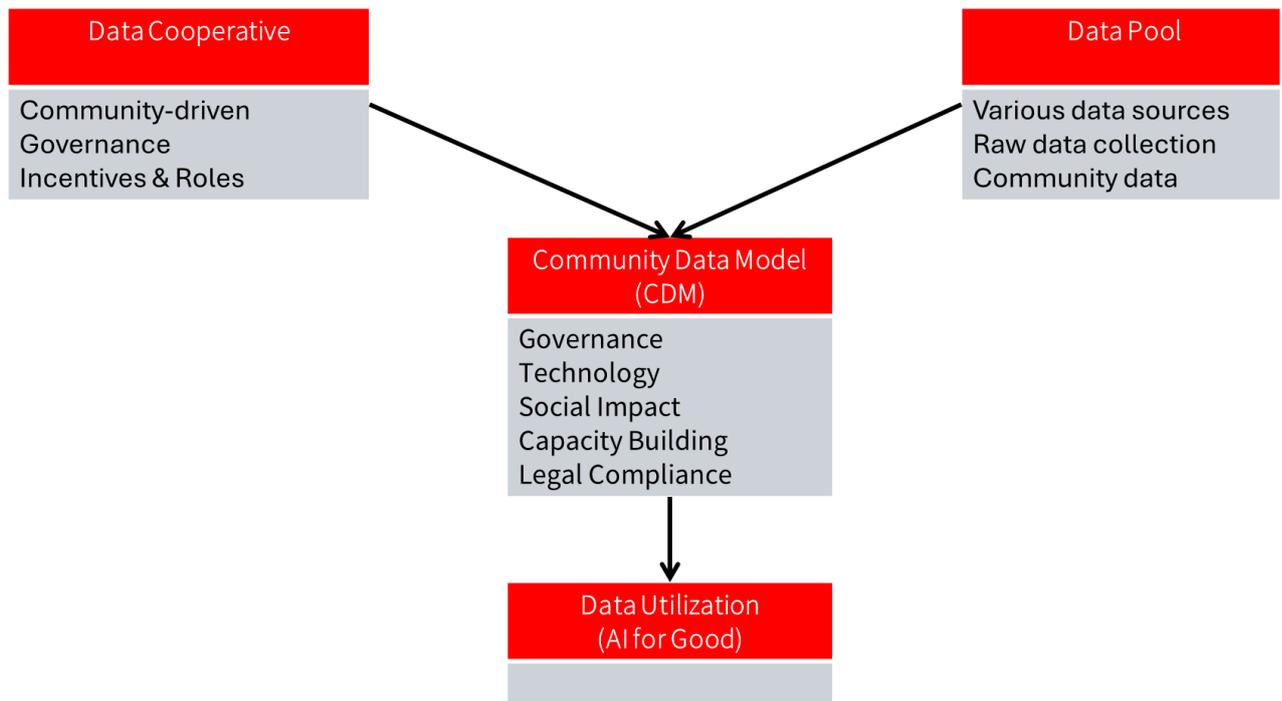

Figure 1: Combining Data Cooperative and Data Pool in the Community Data Model

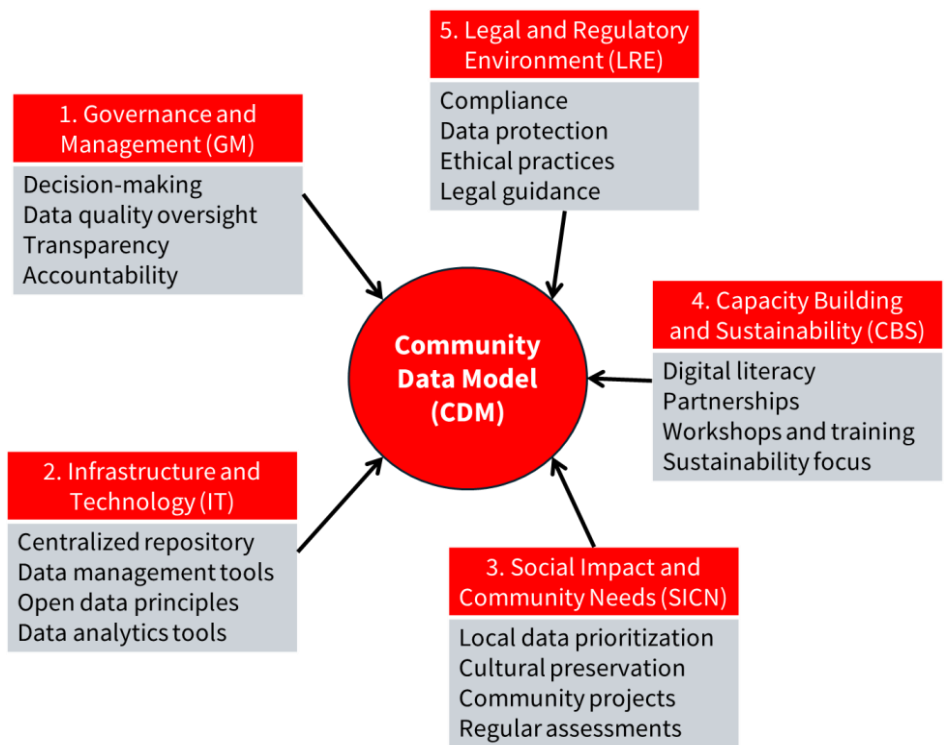

Figure 2: Community Data Model in nutshell.

## 5.1 Governance and management (GM)

The Community Data Model incorporates the governance structure of CNs, which are typically organized as cooperatives. It adds experts to the cooperative board to assist with data management. The board defines the decision-making processes, contributor criteria, and conflict-resolution mechanisms related to the data. Strategies for incentives, such as free Internet access vouchers for contributors or involving students from local schools in data creation process as part of their assignments, can be developed. Board members oversee data quality, ensuring that the collected data are accurate, reliable, and relevant. The cooperative governance model fosters transparency, accountability, and community involvement, ensuring that all stakeholders have voice in the data management process. Regular meetings and feedback mechanisms are defined to allow for the continuous improvement and adaptation of the governance structure.

## 5.2 Infrastructure and technology (IT)

The core of the IT dimension is a centralized data repository capable of handling various community-relevant data types, leveraging existing servers from CNs to host the collected data. The focus is on a data management platform that provides tools for data management, publication, catalogue creation, and facilitates community-wide data sharing. Special attention is given to information lifecycle management for archiving less relevant data, utilizing online archives, such as the Internet Archive[1] or Whose Knowledge[2] platform. The platform upholds open data principles, ensuring accessibility through open formats, free access, clear documentation, usability, and transparency. The platform enables data storage, retrieval, and sharing for the entire community. The platform also integrates data analytics tools, allowing the community to derive actionable insights from collected data.

## 5.3 Social impact and community needs (SICN)

This dimension aims to prioritize local data and knowledge based on local needs and challenges, including educational resources in local languages, cultural knowledge, traditional medicine, etc. An iterative approach, starting with a pilot project that focused on specific domains, is employed. Regular assessments of CDM's impact and identification of areas for improvement are conducted, with the board creating channels for member feedback. This ensures that the CDM remains responsive to the evolving needs of the community, continually improving and adapting to maximize its social impact. By focusing on local data, the CDM helps preserve cultural heritage and traditional knowledge, ensuring that these valuable resources are not lost in the digital age.

## 5.4 Capacity building and sustainability (CBS)

This dimension includes digital literacy training to empower local populations to use smartphones for data collection, curation, documentation and uploading. Schools of CNs can play a significant role in this aspect, as well as educational institutions and organizations that have expertise in data collection. These partnerships help build local capacity, ensuring that community members have the skills and knowledge needed to actively participate in and sustain the CDM. Training sessions are regularly organized to keep the community updated on best practices and new technologies. By focusing on sustainability, the CDM ensures that the community can maintain and grow its data resources over time for a just AI ecosystem.

---

[1] http://web.archive.org/

[2] https://whoseknowledge.org/

## 5.5 Legal and regulatory environment (LRE)

This dimension ensures compliance with national and local data regulations and emphasizes privacy. This includes adhering to data protection laws and ensuring that data sharing practices are ethical and legal. By prioritizing legal and regulatory compliance, the CDM protects the rights of community members and maintains trust in the data management process. Legal experts and advisors are included in the governance structure to provide guidance on compliance issues and to help navigate the complex legal landscape. Additionally, the CDM promotes awareness of data rights and privacy among community members.

# 6 Pillars of data justice in the CDM

The Community Data Model contributes to the pillars of data justice by empowering CNs with control over their data, improving access through open data, increasing participation from diverse stakeholders, ensuring equity for community well-being, and sustaining identity and knowledge by defining data collection priorities and ensuring data quality.

## 6.1 Power

By empowering CNs with control over their data, the CDM allows communities to dictate how their data are collected, managed, and used. This control ensures that data practices align with local values and priorities, giving communities the power to safeguard their own interests and promote their development goals.

## 6.2 Access

The Community Data Model improves access through open data principles, making data readily available to all community members, fostering transparency, and democratizing information. This open access enables researchers, local businesses, and individuals to utilize data for various purposes, from academic studies to economic development initiatives, thus driving innovation and informed decision-making at all levels of the community.

## 6.3 Participation

Increasing participation from diverse stakeholders is a foundation of CDM, ensuring that voices from all segments of the community are heard and considered. This inclusive approach helps build trust and encourages active involvement in data collection and usage, creating a sense of ownership and collective responsibility among community members.

## 6.4 Equity

Ensuring equity for community well-being is achieved by focusing on data that addresses the specific needs and challenges of the community. By prioritizing data that can improve health, education, and economic opportunities, the CDM works to enhance the overall quality of life of all community members, particularly those who are most vulnerable or marginalized.

## 6.5 Identity and knowledge

Sustaining identity and knowledge is a key objective of the CDM. By defining data collection priorities that focus on preserving cultural heritage, traditional knowledge, and local languages, the CDM helps protect and promote the unique identity of the community. This not only enriches the cultural fabric of the community, but also ensures that valuable traditional practices and knowledge are preserved for future generations.

# 7 Leveraging the CDM for AI for good in rural areas

Data collected through the CDM can be leveraged for AI for good in rural areas in some ways.

## 7.1 Local languages and automatic speech recognition systems

As mentioned in (De Vries et al., 2014), mother-tongue speakers of under-resourced languages mostly reside in remote rural areas. The CDM can collect and preserve data in local languages, including audio recordings, texts, and linguistic nuances. AI can use this data to develop Automatic Speech Recognition systems that understand and process local languages. This not only aids in preserving linguistic heritage but also facilitates communication and access to information in native languages.

## 7.2 Agriculture and environmental monitoring

By leveraging data on local farming practices, weather patterns, and soil conditions, AI can provide farmers with personalized advice regarding crop management. This can optimize crop selection and enhance agricultural productivity and sustainability. Furthermore, the utilization of CDM can aid in documenting traditional farming methods that can be integrated with contemporary techniques to improve crop yields, as presented in (J. Fendji et al., 2020) or provide data to feed crop rotation systems like (J. L. E. K. Fendji et al., 2021). AI can use environmental data from the CDM to monitor natural resources, track environmental changes, and predict natural disasters.

## 7.3 Education and healthcare improvement

AI systems can use data from the CDM to create personalized learning experiences for students, addressing the specific educational needs of the community. In addition, AI can analyse health data collected through the CDM to predict health trends and suggest preventive measures. This can lead to more effective health interventions and better resource allocation for rural health services. Moreover, local data can contribute to the integration of AI and traditional medicine in drug discovery (Khan et al., 2021).

# 8 Conclusion

This study underscores the critical shift from traditional CNs to a CDM in rural areas, integrating data pools and data cooperatives to achieve data justice. By leveraging CNs, the CDM establishes a centralized data repository that adheres to open data principles, ensuring accessibility and transparency. The model incorporates cooperative governance structures, prioritizing local knowledge, education, and cultural heritage while also focusing on capacity building. Legal and regulatory compliance are of great importance in this model to ensure data privacy and protection. The CDM not only empowers rural communities by giving them control over their data, but also fosters equitable access and participation, thus sustaining local identity and knowledge. This approach mitigates the challenges of data creation in rural areas, enhances data justice, and significantly improves the quality and relevance of data, thereby enabling these communities to benefit from advancements for a just AI ecosystem. Ultimately, the CDM offers a comprehensive and sustainable framework for data management that supports the long-term development and digital inclusion of rural areas. Future work will aim to design the central platform of the Infrastructure and Technology dimensions, define protocols, and ensure a user-friendly interface for low digital literacy levels.